# Implementation of ISO 23247 for digital twins of production systems:

## Towards AI-based Sustainable and XR-based human-centric manufacturing

**Huizhong Cao[1], Henrik Söderlund[2], Qi Fang[3], Siyuan Chen[4], Lejla Erdal[5], Ammar Gubartalla[6], Paulo Victor Lopes[7], Guodong Shao[8], Per Lonnehed[9], Henri Putto[10], Abbe Ahmed[11], Sven Ekered[12], Björn Johansson[13]**

## ABSTRACT

Since the introduction of Industry 4.0, digital twin technology has significantly evolved, laying the groundwork for a transition toward Industry 5.0 principles centered on human-centricity, sustainability, and resilience. Through digital twins, real-time connected production systems are anticipated to be more efficient, resilient, and sustainable, facilitating communication and connectivity between digital and physical systems. However, environmental performance and integration with virtual reality (VR) and artificial intelligence (AI) of such systems remain challenging. Further exploration of digital twin technologies is needed to validate the real-world impact and benefits. This paper investigates these challenges by implementing a real-time digital twin based on the ISO 23247 standard, connecting the physical factory and simulation software with VR capabilities. This digital twin system provides cognitive assistance and a user-friendly interface for operators, thereby improving cognitive ergonomics. The connection of the Internet of Things (IoT) platform allows the digital twin to have real-time bidirectional communication, collaboration, monitoring, and assistance. A lab-scale drone factory was used as the digital twin application to test and evaluate the ISO 23247 standard and its potential benefits.

Additionally, AI integration and environmental performance Key Performance Indicators (KPIs) have been considered as the next stages in improving VR-integrated digital twins. This paper further discusses how digital twins contribute to the improvement of the environmental performance of production systems, primarily from the perspective of data inventory management. With a solid theoretical foundation and a demonstration of the VR-integrated

[1] Dept. of Industrial and Materials Science, Chalmers University, Gothenburg, SE-41279, Sweden
[2] Dept. of Industrial and Materials Science, Chalmers University, Gothenburg, SE-41279, Sweden
[3] Dept. of Industrial and Materials Science, Chalmers University, Gothenburg, SE-41279, Sweden
[4] Dept. of Industrial and Materials Science, Chalmers University, Gothenburg, SE-41279, Sweden
[5] Dept. of Industrial and Materials Science, Chalmers University, Gothenburg, SE-41279, Sweden
[6] Dept. of Industrial and Materials Science, Chalmers University, Gothenburg, SE-41279, Sweden
[7] Dept. Computer Science, Aeronautics Institute of Technology, São José dos Campos - SP, 12228-900, Brazil
[8] NIST, Gaithersburg, Maryland, US
[9] PTC, Gothenburg, Sweden
[10] Rockwell Automation, Gothenburg, Sweden
[11] Rockwell Automation, Gothenburg, Sweden
[12] Dept. of Industrial and Materials Science, Chalmers University, Gothenburg, SE-41279, Sweden
[13] Dept. of Industrial and Materials Science, Chalmers University, Gothenburg, SE-41279, Sweden

digital twins, this paper addresses integration issues between various technologies and advances the framework of digital twins based on ISO 23247.

**Keywords**

Digital twins, ISO 23247, AI, XR, Sustainability, Real-time connectivity, Industry 5.0

# 1 Introduction

In recent years, as industries have rapidly adopted Industry 4.0 technologies to process and leverage real-time data, attention is now turning toward Industry 5.0 principles that emphasize human-centricity, sustainability, and resilience (Breque et al., 2021; Neumann et al., 2020b). While Digital Twins (DTs), the virtual counterparts of physical assets, have long enabled real-time monitoring, predictive maintenance, and optimization, their role is now expanding beyond the seamless integration of digital and physical domains (Perno et al., 2022; Ariansyah et al., 2020). In line with the European Commission's Industry 5.0 vision, DTs are increasingly combined with Extended Reality (XR) and AI-driven sustainability key performance indicators, creating production systems that can adapt more swiftly, respond more effectively, and enhance user experiences, well-being, and environmental responsibility (Cao et al., 2024). By employing these immersive XR capabilities and advanced AI metrics, it becomes possible to interact intuitively with digital counterparts, ensuring that human values guide the evolution of resilient, sustainable, and adaptable manufacturing ecosystems (Schuh et al., 2020). Moreover, using DTs for human modeling and simulation enables ergonomic assessment during operation, which helps reduce physical and cognitive loads by predicting and optimizing workers' motions and behaviors (Maruyama et al., 2021).

Incorporating cognitive ergonomics into a DT environment through an XR interface enhances the usability and effectiveness of these systems by optimizing how users perceive, monitor, and interact with complex information provided by the system (Asad et al., 2023; Greco et al., 2020). DT and virtual reality technologies can be combined for designing, simulating, and optimizing cyber-physical production systems, and they can interact remotely or collaboratively (Havard et al., 2019).

The purpose of the study is as follows:

- To establish a use case for the ISO 23247 standard (Shao and Helu, 2020) in a laboratory setting, emphasizing real-time connection in a practical setting to link the virtual and physical drone factories.
- To further investigate the real-time connectivity use case for applying cognitive ergonomic principles through XR user interfaces, which might enhance user engagement and decision-making in DT systems
- To bridge the gap between DT's framework for Industry 5.0, which is human-centric, resilient, and sustainable manufacturing

- To demonstrate the journey from the straightforward use case to sustainability initiatives, AI integrations, and virtual manufacturing with cognitive ergonomics.

The main research questions in this study are:

- RQ1: *How can the ISO 23247 standards be used to support the integration of simulation models and an IoT platform, establishing real-time connectivity?*

- RQ2: *How can the ISO 23247 standards be implemented in a Digital Twin lab setting, integrated with sustainability, AI, and XR technologies for Industry 5.0?*

To answer these research questions, this study provides a practical real-time DT use case implementation using the ISO 23247 framework. An integrated DT reference architecture based on the ISO 23247 standard is created, a VR-integrated simulation model is developed to represent a lab-scale real system and an architecture is established to support the connectivity between the simulation model and the real system. Section 2 of the study outlines the relationship between DT, VR, and IoT technologies. Section 3 introduces the methodology, which is based on the standard framework for DT development and how it is applied in real-world use case scenarios. Section 4 presents and discusses the findings, including a practical case study of a lab-scale drone factory monitored and controlled in real time. Section 5 concludes the paper by proving the value and benefits of incorporating Extended Reality (XR) and AI into the DT platform, as well as sustainability performance in the future.

# 2 Theoretical Background

This paper begins with a theoretical foundation on the three interrelated themes to position the study within a broader context. The subjects covered include DTs and their functions, including advantages and various implementation views; VR and simulation, including terminology and applications; and IoT systems, including components and setups.

## 2.1 DIGITAL TWINS AND SIMULATION

The concept of DTs is a dynamic mapping between physical items and simulation models, incorporating the physical and cyber levels of a system (Grieves, 2014). This system promotes smart manufacturing processes by enabling connectivity, interaction, control, and management on the shop floor (Tao and Zhang, 2017). One distinct advantage of DTs is their ability to identify systemic difficulties in complex systems, which are frequently caused by human interactions during operations (Grieves and Vickers, 2017). Furthermore, DT implementations allow data-driven and smart production, laying the groundwork for future Industry 4.0 technologies (Qi and Tao, 2018).

The literature on DTs focuses on diverse viewpoints of DT implementations, from cataloging system architecture patterns in DT designs (Tekinerdogan and Verdouw, 2020) to outlining the scope and structural design requirements for DTs (Shao and Helu, 2020b). The ISO 23247 standard for DTs addresses interdisciplinary implementation challenges by providing a framework for manufacturing DT implementations (Shao et al., 2023).

Jiang et al. (2021) give a comprehensive analysis of DT applications in manufacturing, highlighting considerable operational efficiencies in process monitoring, life prediction, and asset management (Jiang et al., 2021). There are more examples of ISO 23247 implementations, including modeling DTs for flexible manufacturing cells, stressing critical characteristics, and product lifecycle integration (Wallner et al., 2023). Kim et al. (2022) describe the architecture of a wire arc additive manufacturing DT. Although not completely implemented, the advantages were rated as having a clear potential to promote integration and real-time decision-making.

DTs support decision-making, improve visibility, optimize operations, and strengthen resilience to disturbances (Lugaresi et al., 2023). To ensure confidence and dependability in manufacturing DTs, it is crucial to focus on verification, validation, and continual credibility maintenance (Shao et al., 2023). Integrating immersive VR with DTs can increase visualization and human interaction while overcoming hurdles like complicated VR design, hardware controls, and communication methods (Pirker et al., 2022). These findings demonstrate the revolutionary potential of DTs in numerous areas, solving difficulties in areas such as harmonizing communication interfaces and missing efficient modeling frameworks (Sun et al., 2022).

Simulation and VR are key technologies revolutionizing various industries, particularly for manufacturing. Simulation involves creating a digital representation of a real system or process to analyze its behavior or performance. It is instrumental in optimizing production systems, as it allows for scenario testing and decision-making based on predicted outcomes (Ottogalli et al., 2019). VR, on the other hand, is an interactive computer-generated experience that replicates settings or scenarios, frequently utilizing immersive technology.

## 2.2 EXTENDED REALITY

XR technologies are frequently used to define a range of diverse technologies that can serve as a bridge between the virtual and real worlds. A spectrum is known as the virtual continuum (Milgram and Kishino, 1994). The XR technologies on the virtual continuum offer different levels of virtual immersion through display technology, with one end of the spectrum offering a fully virtual experience while the other end offers a fully physical experience (Milgram and Kishino 1994). XR technologies in industrial training programs can create immersive, personalized experiences while considering user and workplace specificities, enhancing ergonomics and knowledge retention (Pavlou et al., 2021). XR systems can improve efficiency and ergonomics in industrial environments, particularly for workers with disabilities, by automating content creation and facilitating skills transfer (Simões et al., 2019).

In broad terms, these technologies are often simplified as VR (virtual reality), MR (Mixed Reality), and AR (Augmented Reality), see Figure 1.

FIGURE 1 THE VIRTUAL CONTINUUM, ADAPTED FROM (MILGRAM AND KISHINO 1994)

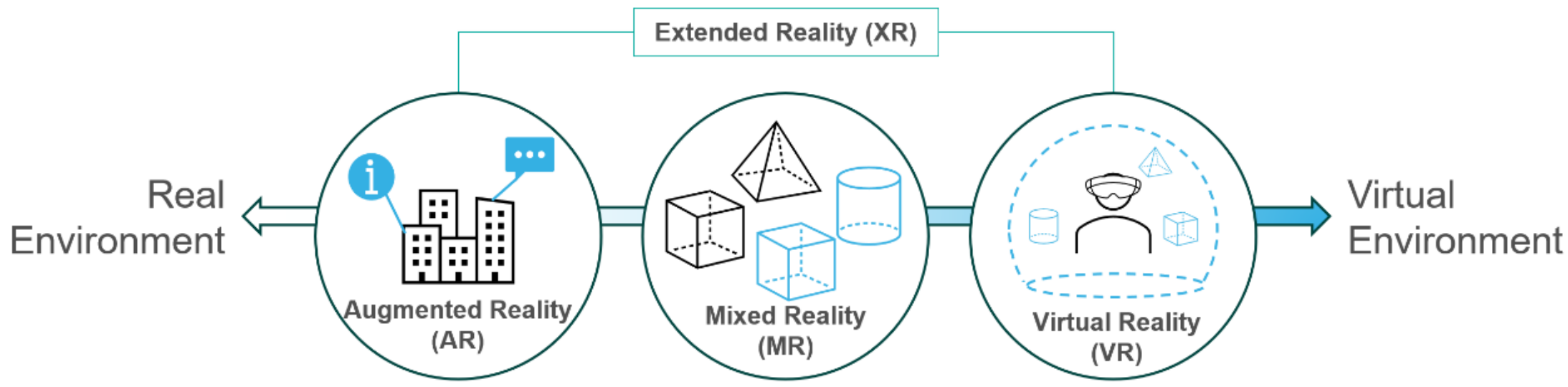

AR technologies enhance the physical world by integrating virtual enhancements with real-world objects (Saju et al., 2022). Most notably, this is usually done using AR glasses or smart glasses but could also be achieved using mobile devices and displays. AR technologies combine real and synthetic objects by superimposing computer-generated annotations on the real environment, enhancing the user's perception and helping them perform specific tasks (Meike van Lierop et al., 2021). Using a DT of the environment, users can interact more efficiently even with a limited field of view (FOV) using AR glasses while maintaining spatial understanding (Shin & Kim, 2022).

MR technologies seamlessly mix the physical and the virtual worlds by blending virtual and physical objects and assets and enabling interaction. MR technologies rely on both a physical system or asset and a virtual world that can dynamically adapt and synchronize with the physical world (Farshid et al., 2018). MR environments, when integrated with DTs, can reduce the mental and physical workload of operators during teleoperation, enhancing the human-robot interaction experience and teleoperation performance (Fan et al., 2023). One case is that Digital Twin Mixed Reality (DTMR) architecture combines DT real-time mapping in MR environments for realistic asset behavior and ergonomic operators' characteristics evaluation (Xia et al., 2022).

VR technology is the most commonly discussed technology among the three and solely relies on the virtual world. The connection to physical assets is very low or non-existent, implying that interactions within the virtual world are possible (Farshid et al., 2018). VR and Digital Human Modeling can be combined to assess physical ergonomics, potentially improving operators' well-being and productivity by detecting ergonomic issues early in the product development process (Da Silva et al., 2022).

Given the context of DTs, the term proposes the existence of both a virtual and a physical system. In such a context, all the technologies from the virtual continuum could find relevant use cases, depending on which part of the DTs the user is interacting with. Users interacting with the physical system or asset could benefit from AR or MR technology to augment the physical system with information or data from the DTs. On the other hand, users interacting with only the virtual part of the DTs could benefit from using VR as the interface to the system or asset.

### 2.3 IOT SYSTEMS

An IoT system connects the actual system to the simulation model in a DT via linked devices such as sensors and actuators. Applications employ collected data for a variety of purposes, including predictive maintenance, analysis, monitoring, control, and optimization (Guth et al., 2018). However, one major problem with DT frameworks is developing efficient communication between virtual models and actual systems. To enable synchronous interactions between a simulation model and its physical counterpart, communication protocols such as OPC UA, CODESYS, and RESTful APIs/MQTT are used (Guth et al., 2016). Technological problems stem from a lack of domain experience and understanding of communication and information technology (Chen et al., 2024).

To describe an IoT system, there exists a number of IoT architecture frameworks provided by both the research and manufacturing industry. (IEEE 2020) defines a comprehensive framework standard including the architecture development process and application domains, the 3- or 5-layered models bring a more simplified architecture solution (Guth et al., 2016), while (Swamy and Kota 2020) provides a hybrid architecture with examples of the various components of the IoT system.

The architecture used for describing the different components is based on the IoT reference architecture proposed by (Domínguez-Bolaño et al., 2022), which addresses crucial components adopted in this paper. It comprises sensors, actuators, the IoT platform, a communications network, an IoT gateway, applications, security, and management. The sensors capture information supplied to the user to control the system by using the actuators, which convert the command to a physical action based on a request from the IoT platform. IoT platforms, gateway components, and communication networks based on various communication standards provide communication between sensors and actuators. The elements supporting the communications are the middleware component IoT gateway, the data layer protocol, and the prescribed data format. The IoT platform is responsible for managing the devices, handling data storage and analytics, and supporting applications, where the software uses the input and output from devices to perform tasks and services. Security is maintained by encrypting and monitoring keys and certificates, whereas management regards system performance and configuration, networks, fault management, and service updates. To facilitate successful real-time connectivity with these components, a methodology for the integration is required to solve application gaps such as standardization, interoperability, data acquisition, and poorly constructed user interfaces (Chen et al., 2024), which is presented in the following section.

### 2.4 THE ROLE OF AI IN DTS

AI is essential for unlocking the full potential of DT (Emmert-Streib 2023), transforming them into dynamic and intelligent systems that mirror their physical counterparts in real time. By integrating AI, DTs can go beyond simple replication to become powerful tools for predictive maintenance, optimization, and decision-making. The synergy between AI and DT increases efficiency, reduces downtime, and enables smarter, data-driven operations (Jazdi et al., 2021). AI's role is multifaceted, enhancing the ability to collect data from physical systems, improve the accuracy of DT models, and enable real-time updates (Rossini et al., 2020). AI also strengthens the analytical and simulation capabilities of DTs, allowing for better optimization (Zhang et al., 2020). Furthermore, AI empowers DT to make autonomous, data-driven decisions (Shen et al., 2023).

Updating DT in real-time presents a significant challenge, but AI plays a crucial role in overcoming it. AI can be used to automatically collect, pre-process, cleanse, and integrate data from multiple sources, allowing higher quality data to be used to create or maintain DTs, ensuring that they are accurate and reflect the current state of the physical system in real-time. IoT-based AI systems, such as the IBM's Watson IoT platform, which use machine learning algorithms to gather data from multiple connected sensors and devices (Cloud 2024). These systems autonomously collect data from the physical environment in real time, ensuring a continuous source of data for DTs. Google's TensorFlow can be used for data preprocessing, help normalize, scale, and transform raw data into a format beneficial to input in DTs, filtering out unnecessary or irrelevant data points before analysis (TensorFlow 2024).

## 3 Methods

The purposes of this study are (1) to establish a use case for the ISO 23247 standard (Shao and Helu, 2020) in a laboratory setting that emphasizes real-time connectivity between virtual and physical drone factories, (2) to investigate how cognitive ergonomic principles, applied through XR-based user interfaces, can enhance user engagement and decision-making in DT systems, (3) to bridge the gap between the human-centric, resilient, and sustainable manufacturing vision of Industry 5.0 and current DT frameworks, and, (4) to demonstrate a progression from an initial, straightforward use case toward sustainability initiatives, AI integrations, and virtual manufacturing supported by cognitive ergonomics. By applying these approaches, the study aims to design, test, analyze, and optimize manufacturing processes within a virtual environment. The following subsections introduce the physical system and its components, which have been developed and tested as a proof-of-concept to validate the proposed framework and incorporate the practical drone factory use case. The demonstration of real-time connectivity will serve as a basis for subsequent discussions on AI-enhanced digital twins with XR interactions and sustainability as a key performance objective.

### 3.1 PHYSICAL SYSTEM

The physical system is located in Lindholmen, Gothenburg, at the Stena Industry Innovation (SII) Laboratory. This lab provides a platform for smart industrial production research, training, and demonstration within the framework of digital transformation and Industry 4.0. It offers a chance to use simulation models, an Internet of Things (IoT) platform, and embedded sensors to verify real-time connectivity.

The establishment has a lab-scale drone factory with a production line that transfers drones on pallets via a conveyor belt. The real system and its simulation model are shown side by side in Figure 2. Six branch assembly stations, including an Autonomous Mobile Robot (AMR) and a buffer space, are connected by a conveyor belt. To oversee the conversion of products and raw materials both internally and externally via the main conveyor, the lab also uses internal logistics (Taylor et al., 2021).

Programmable Logic Controllers (PLCs) control the lab-scale drone manufacturing, and each station has input/output (I/O) units. Pallets with Radio Frequency Identification (RFID) tags attached are detected by sensors at each station. Pallets can travel vertically between the main conveyor and the workstations; thanks to the docking station's I/O arrangement, which has two sensors for pallet detection and two for elevator transfer monitoring. Furthermore, a sensor controls the queue stop to regulate pallet flow (Stena Industry Innovation Laboratory, 2024).

**FIGURE 2 REAL AND VIRTUAL SYSTEMS OF THE SMART PRODUCTION SYSTEM**

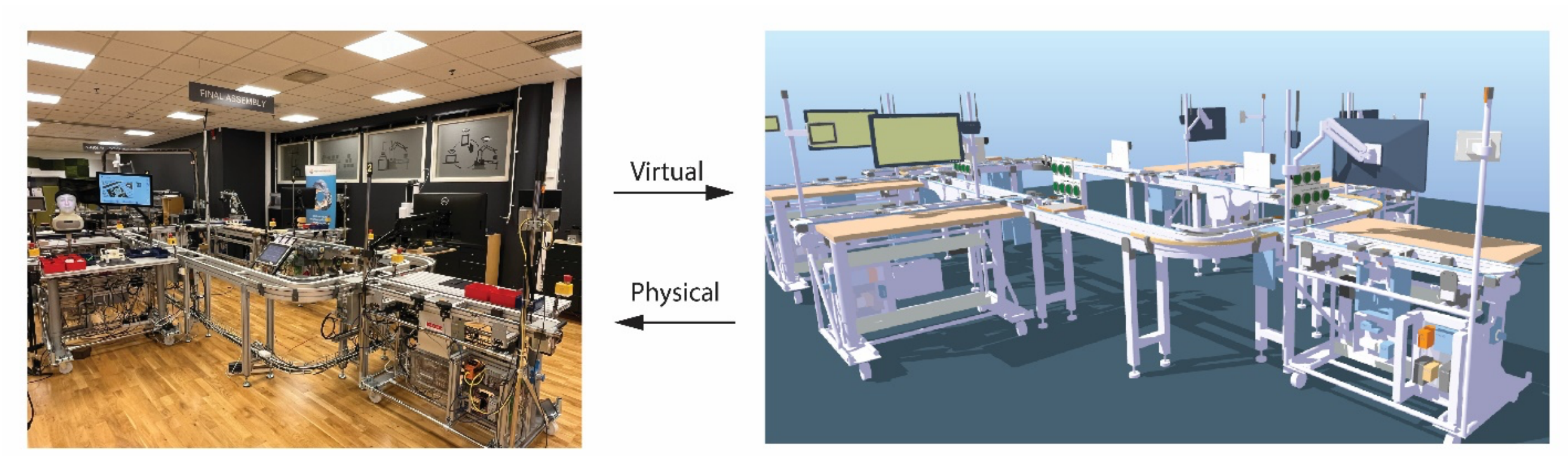

(a) SII Lab - real system (b) Emulate3D - simulation model

## 3.2 ISO 23247, THE STANDARD FRAMEWORK FOR DEVELOPMENT OF DTS

This section outlines a methodology based on the ISO 23247 standard for implementing DTs with real-time communication. For human-centered interaction, this framework facilitates smooth communication between the simulation model, the physical system, and virtual reality (VR) (Shao et al., 2021). Interoperability and standardization are made easier by adopting ISO 23247, which guarantees that the DT implementation complies with globally accepted best practices. ISO 23247 comprises four main components:

1. General Principles: Define the physical systems as Observable Manufacturing Elements (OMEs) that need to be twinned.

2. Reference Architecture: The standard outlines a reference architecture consisting of four main domains:
    - OME Domain: Represents the physical manufacturing elements and their behaviors.
    - Data Collection and Device Control Domain: Manages the acquisition of data from the OMEs and controls devices as necessary.
    - Core (DT) Domain: Contains the digital representations and models of the OMEs.
    - User Domain: Provides interfaces for users to interact with the DT, including visualization and analytics tools.
3. Digital Representation: This component describes the essential information attributes for typical OMEs, encompassing both static information (e.g., design specifications) and dynamic information (e.g., real-time sensor data).
4. Information Exchange: It outlines the technical requirements for exchanging information between entities within the framework, ensuring consistent communication protocols and data formats.

These components provide guidelines for defining the scope and objectives, setting simulation model requirements, implementing a generic reference architecture, and supporting information synchronization between the DT and its physical counterpart.

The standard reference architecture is instantiated in the next section for the implemented use case, demonstrating how it can be applied to a practical scenario. In addition to ISO 23247, existing standards developed for specific areas—such as data collection, data governance, data security, information modeling, system modeling, simulation, visualization, and networking—can be leveraged to support the development of DT applications (Shao et al., 2021). Utilizing these standards enhances the robustness and scalability of the DT framework, as shown in Figure 3 and expanded in Figure 7.

Furthermore, the Asset Administration Shell (AAS) is another DT implementation standard that can be considered for future use cases (George & Henry, 2017). The AAS provides a standardized digital representation of assets, facilitating integration and communication within Industry 4.0 environments.

## 3.3 VIRTUAL REALITY INTEGRATED SIMULATION

The simulation model of the lab-scale drone factory was built using Emulate3D, a simulation, and DT platform software. A library of pre-modeled components and built-in catalogs with manufacturing machinery like walls, staircases, conveyors, robots, and loads are included in this software. Computer-Aided Design (CAD) files for the drone product are imported, and the software automatically incorporates the required features. Users can test connectivity in VR utilizing the High-Tech Computer Corporation (HTC) Vive VR platform for human-centric interaction by combining the virtual model with features like sensors and buttons. The virtual model can be used for simulation, emulation, or demonstration.

In addition, the simulation tool supports a number of commercial off-the-shelf (COTS) and standardized communication protocols, including CODESYS and OPC UA. Tags are used to indicate the input/output (I/O) points of the physical system. The IO Browser defines these tags, shows the tags that are currently loaded, together with their definitions and bindings, and links them to the controller using a Tag Server. It is advised to eliminate superfluous physical attributes and use simplified representations of real objects to guarantee the simulation's efficacy.

## 3.4 INDUSTRIAL INTERNET OF THINGS ECOSYSTEM AND COMMUNICATION MIDDLEWARE

For the lab-scale drone factory, we used PTC ThingWorx to facilitate the development and administration of the Industrial Internet of Things (IIoT) ecosystem. This platform facilitates a modular approach to data modeling while offering scalability and practical connectivity. ThingWorx's Foundation Server is used for data collection, storage, and visualization. The "Things," "Thing Templates," and "Thing Shapes" are used to build applications. In ThingWorx, a Thing might be a tangible asset, system, process, product, person, or device. Because each Thing is built on a Thing Template, which specifies a collection of attributes and functions, similar Things may be duplicated and configured consistently. The Thing Shape uses attributes, events, subscriptions, and services to describe the relationships between things. Scalable platform maintenance and updates are made possible by Thing Templates' ability to inherit properties and business logic from Thing Shapes (PTC Community Management, 2022).

To facilitate secure and efficient communication across various levels of the operations network, we employed Kepware as middleware. Kepware allows encrypted communication and manages traffic across multiple firewalls, limiting unauthorized access while permitting remote administration and setup of connected devices within the factory network. Additionally, Kepware can function as an OPC UA (Open Platform Communications Unified Architecture) server, enabling real-time, bidirectional communication between the simulation model and Kepware.

Communication between Kepware and the physical system was established via CODESYS. We made sure that there was smooth data transfer and control between the middleware and the real equipment in the lab-scale drone production by integrating CODESYS.

## 4 Results

Following the completion of objectives with the reference architecture based on ISO 23247, the VR simulation model was presented, the integration of the IoT platform and the simulation model was described, and connectivity was realized with the help of a proof-of-concept use case.

This part answers RQ1: *How can the ISO 23247 standards be used to support the integration of simulation models and an IoT platform, establishing real-time connectivity?*

### 4.1 DEVELOPED FRAMEWORK BASED ON ISO 23247

Figure 3 presents an instantiated functional view of the ISO 23247 standard architecture (Shao 2021) tailored for the lab-scale drone factory DT implementation. As stated in Section 3, the standard framework is divided into four sections, but this study focuses solely on the reference architecture component as in Figure 3, and an extension of the use case based on the reference architecture will be shown in Figure 7.

Each domain in the reference architecture represents a logical set of activities and functions carried out by the functional entities (FEs): OMEs are physical elements from which data are gathered using multiple devices, such as sensors, PLCs, and CAD files. The core DT entity includes and manages simulation modeling, VR modeling, networking, synchronization, interoperability, and other functionalities as described in Section 3. The lab-scale drone factory DT performs real-time monitoring and PLC control; and the user entity lets users communicate with the core entity, VR controllers, and the IoT platform. The reference architecture makes it easier to construct DTs by letting developers select the relevant components for implementing the DT for the use case.

**FIGURE 3 FUNCTIONAL VIEW OF THE INSTANTIATED REFERENCE ARCHITECTURE, ADAPTED FROM (SHAO ET AL., 2021)**

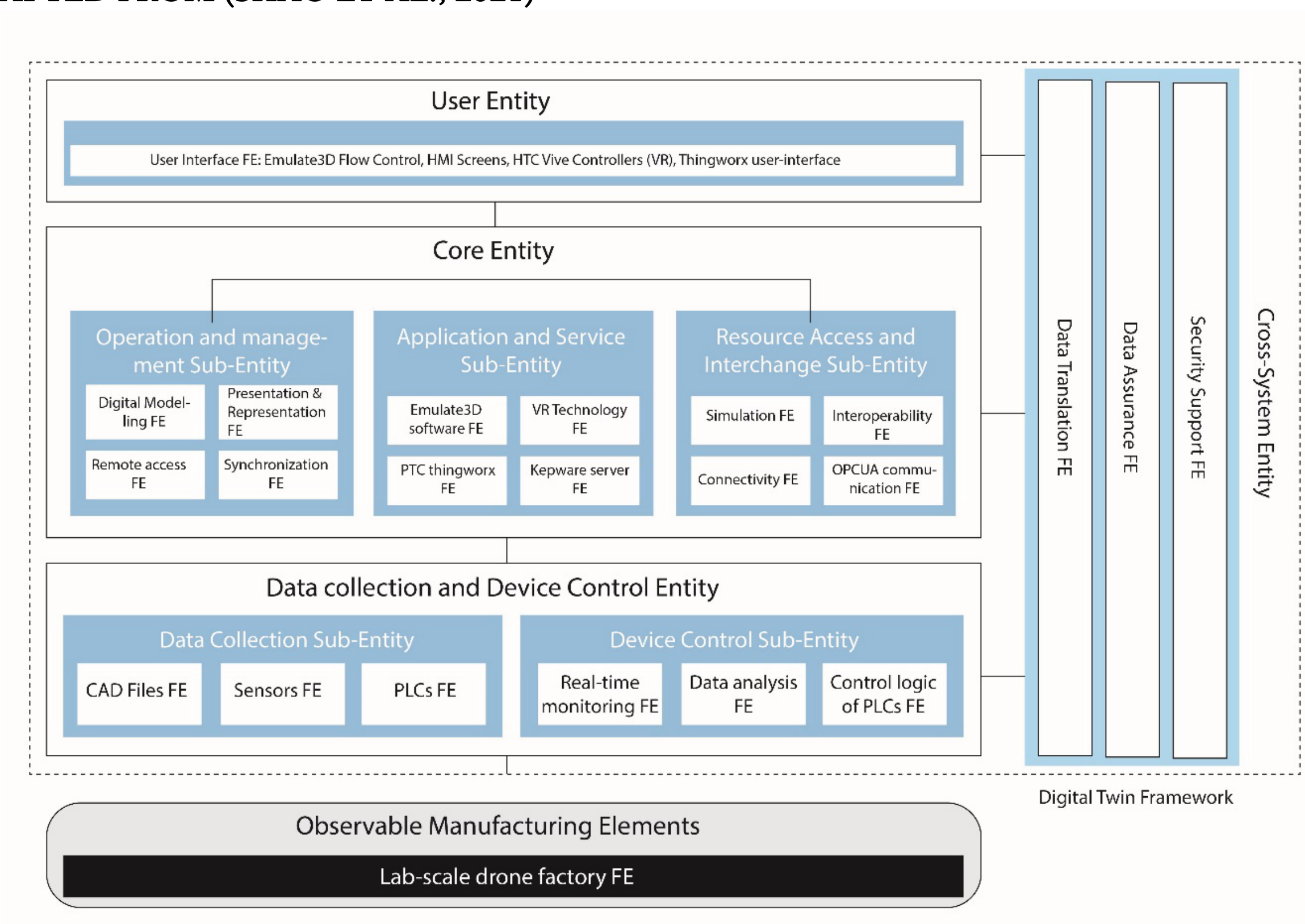

## 4.1 INTEGRATION AND DEVELOPMENT RESULTS

One of Emulate3D's VR features is the ability to transmit the human viewpoint of the virtual system, which forms the basis for applications such as virtual operator commissioning and training. Pallets, drones, and kits function as movable, interactable objects with physical characteristics like gravity and friction on the desktop view, where interactions are pre-edited. Virtual control can be accessed immediately from the VR environment through a headset by pressing a button in both the VR scene and the PLC control system in the real world, as seen in Figure 4.

**FIGURE 4 HUMAN PERSPECTIVE IN VR ENVIRONMENT TO CONTROL THE DT SYSTEM**

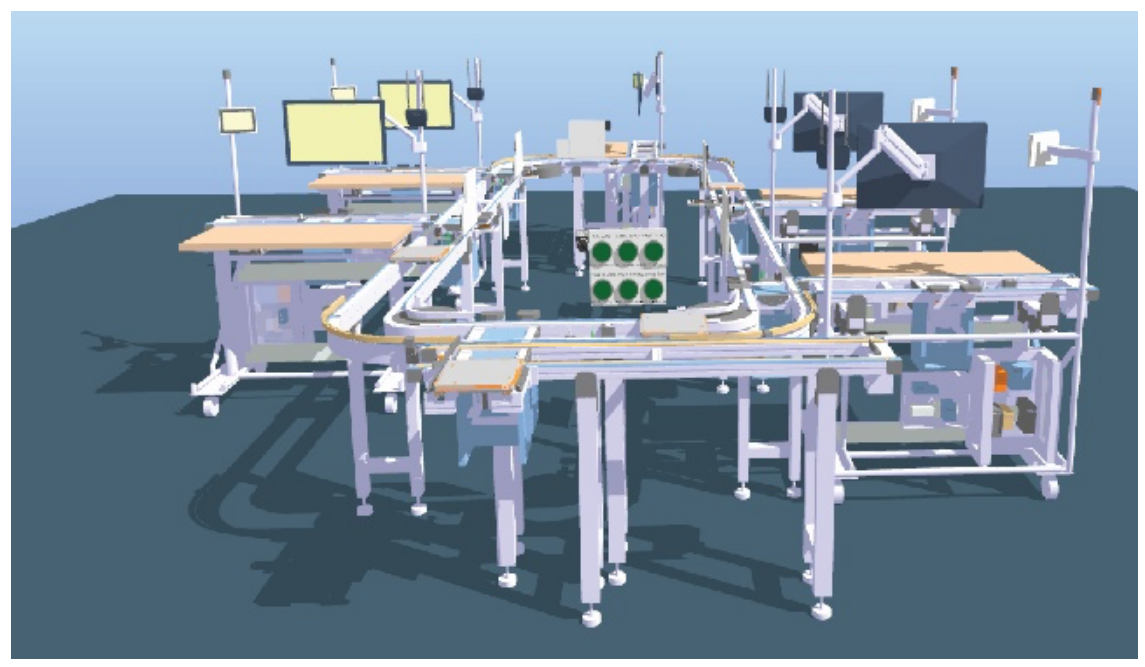

## 4.2 INTEGRATION OF SIMULATION MODEL AND IOT PLATFORM

The connection was maintained through the Kepware server while the simulation models were finalized and the IoT platform was configured. Secure data authentication was ensured via an application key, making connecting and monitoring the lab-scale drone factory's equipment easier. Furthermore, IP address whitelisting was implemented by the server to improve security and manage data access (PTC Community Management, 2023). An OPC UA server was built in Kepware to connect the simulation model to the server. This allowed for a client/server architecture, making data transfers via firewalls safe and strightforward. When the simulation model is coupled with the real system via the IoT platform with Kepware, which represents the DT, Figure 5 illustrates the communication structure made possible by the created connectivity. Real-time VR immersion in the model allows the user to interact with the DT.

**FIGURE 5 COMMUNICATION BETWEEN THE REAL SYSTEM AND SIMULATION SOFTWARE**

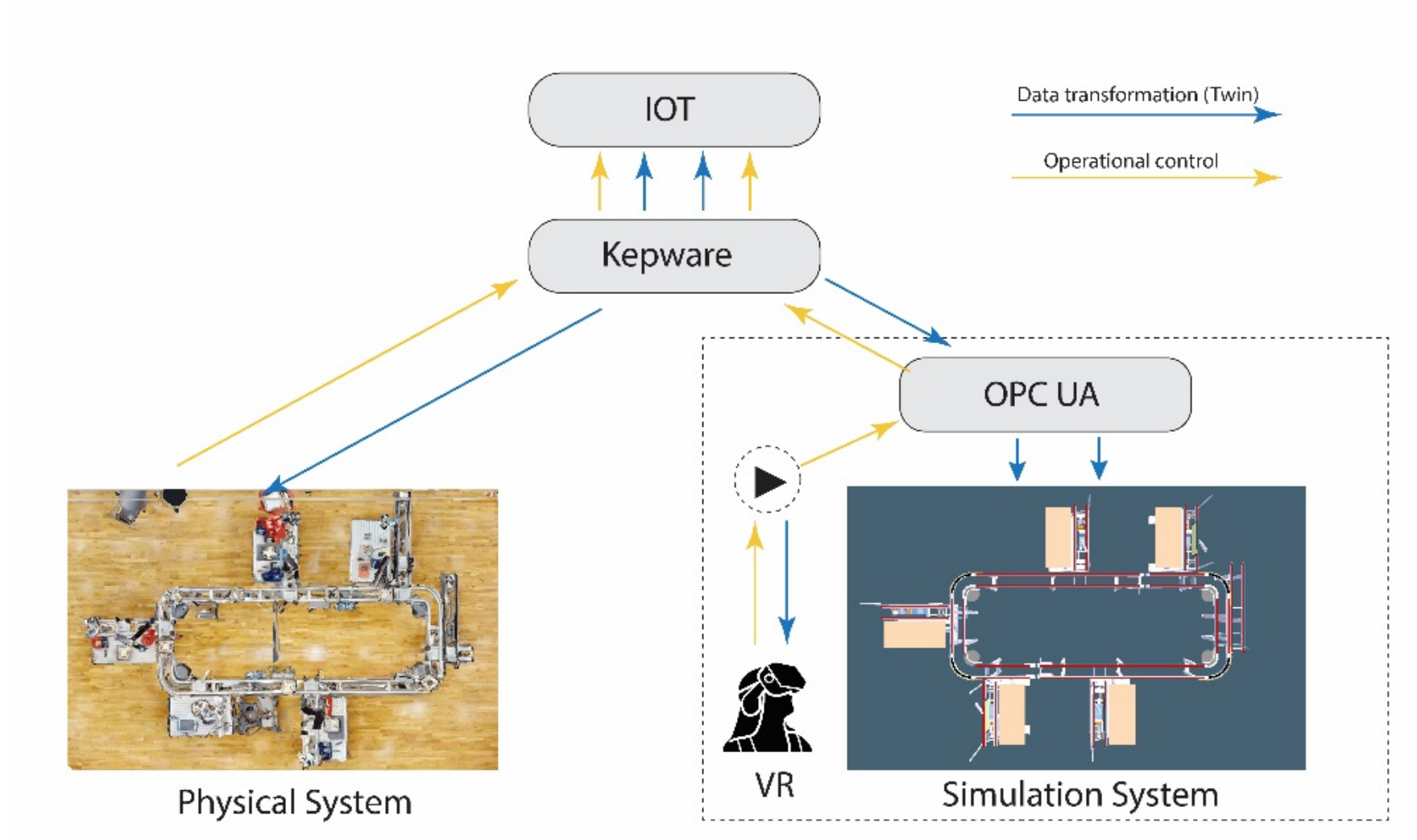

Because the real system is the master, the simulation model must send out a request to start a mission. The OPC UA communication standard is used to send this request through the Kepware server. The request is then forwarded to ThingWorx, which uses CODESYS to forward it to the relevant station in the lab-scale drone factory after confirming the mission's availability. This solution ensures smooth communication and execution by directly interacting with the PLCs connected to the server. Before the task is carried out, the simulation model receives validation from the real system. A delay of roughly one second is achieved in communication between the simulation model and the IoT system with the selected configurations. According to Lopes et al., (2023), this latency impacts real-time data updates, and setups may need to be changed based on the situation. In this use instance, the DT was primarily used for monitoring and control, ensuring synchronization between the simulation model and the real system requires minimal latency.

### 4.3 USE CASE VALIDATION RESULTS

The validation of our DT framework was conducted through a comprehensive process to ensure the system's accuracy, reliability, and effectiveness. This validation is crucial to demonstrating compliance with the ISO 23247 standard and establishing the DT's capability to enhance manufacturing processes through sustainability, AI integration, and virtual production.

The general validation was conducted through a proof-of-concept use case that visually and functionally confirmed the synchronized behavior of the physical system, simulation model, and VR user interactions (as depicted in Figure 6). We utilized multiple methods for initiating system commands—VR controllers, the IoT platform, and the physical human-machine interface (HMI) screens—to test the system's robustness and flexibility. The DT's ability to replicate physical actions in real time and support bidirectional communication validates its operational effectiveness.

FIGURE 6 VISUALIZATION OF THE BI-DIRECTIONAL CONNECTIVITY THROUGH VR

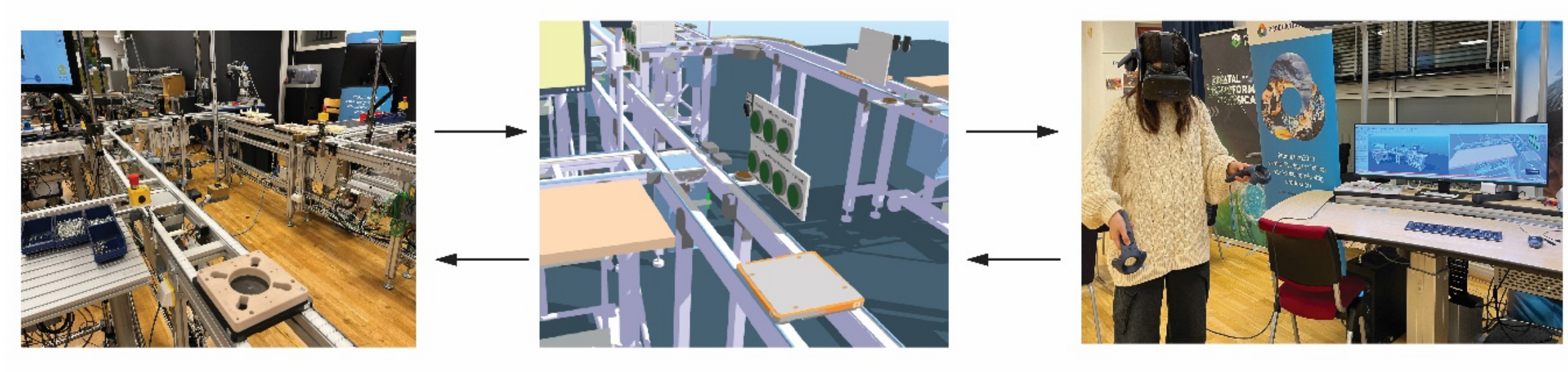

(a) Real System (b) Simulation Model (c) VR-User

**Validation of the Simulation Model**

The simulation model was validated by ensuring its behavior accurately mirrors that of the physical drone factory system. This was achieved by initiating specific missions, such as requesting a drone pallet to pass a docking station and observing the synchronized events between the real system and the simulation. Visual confirmation was made by comparing the actions in the physical system (Figure 6a) with those in the simulation model viewed through VR (Figure 6b) and the VR user's perspective (Figure 6c). The alignment of movements, responses, and system states confirmed the simulation model's validity.

**Validation of Server Connections**

The server connections were validated by establishing robust communication links between the PLC using CODESYS, the IoT platform, and the simulation environment. We tested the bidirectional data transfer by sending commands from the VR interface and the simulation model to the physical system and vice versa. The successful real-time execution of commands, data updates, and system responses across these platforms validated the connectivity and interoperability of the system components as per ISO 23247 guidelines.

**Validation of VR Use Cases with Proper Description**

VR integration was validated when users were able to interact with the DT through immersive VR interfaces. Users could perform actions such as pressing a virtual push button using HTC Vive controllers to initiate missions in the simulation, which were then reflected in the physical system. This interaction was monitored to ensure real-time synchronization and accurate system responses. The VR use cases demonstrated the effectiveness of VR as an interface for controlling and monitoring the DT, enhancing user engagement, and supporting cognitive ergonomics.

Our validation process closely aligns with the ISO 23247 standard, which emphasizes the importance of standardized architectures, digital representations, and information exchange in DT implementations. By ensuring that each component of our DT framework—from the simulation model to server connections and VR interfaces—operates cohesively and communicates effectively, we adhere to the standard's requirements. The successful validation of interoperability, real-time data exchange, and system synchronization showcases our compliance with ISO 23247.

Table 1 provides a detailed description of the use case, where the selected mission is to request the drone pallet to pass a docking station. The sensor that obstructs the passage deactivates, and the pallet can proceed to the next station. To make a request, the user has several options. They can use the headset and controllers to make virtual requests by pressing a connected virtual push button, placing the request using the user interface, or directly communicating with the PLC through the HMI screen. The objective is to ultimately have the simulated pallet mirror the behavior of the physical pallet in terms of both speed and position. However, the RFID tags have not yet been incorporated into the DT, which is recommended for future research to develop a system that updates the pallet's position in real time while adjusting it to the communication latency.

TABLE 1: A SUMMARY OF REAL-TIME CONNECTIVITY USE CASE

| *ID* | SII lab Scale Drone Factory |
|---|---|
| *Use case name* | Connectivity between IoT and simulation software |
| *Application field* | Smart Manufacturing |
| *Cycle stage(s)/phase(s) coverage* | Production |
| *Status* | TRL 7 Demonstration in a representative environment |
| *Scope* | Perform real-time connectivity in DT |
| *Initial (Problem) situation* | Within the domain of DT frameworks, a fundamental challenge lies in establishing connectivity between the virtual and physical systems. In this context, connectivity refers to the communication protocol, enabling synchronous interactions between a virtual model and its physical counterpart. The goal is to facilitate real-time control and movements of the physical entities while ensuring that the DT accurately replicates these actions. The challenge is to address how this data connection operates the simulation-based DT and the physical system ensuring a connection capable of bidirectional data transfer. |
| *Objectives* | Implement the real-time connectivity demonstrator in a physical system |
| *Short description* | To demonstrate the developed connectivity, missions from PLCs have been virtually created in a simulation tool, represented as push buttons. The logic mimics the real system's progressions. The selected mission is Pass Docking Station, where the user can place the request either in the simulation software, at the IoT platform, or directly with the HMI screens belonging to the real system. The blocked drone pallet can then pass the docking station, where the mission is replicated in the simulation model in real time. |
| *Stakeholders* | Manufacturing shop floor personnel, management, researchers |
| *Key technologies* | Automation, data analytics, simulation, VR, IoT |

## 5 Discussion

In this section, we delve into the Drone Factory DTs framework, focusing on three main objectives to answer the RQ2: *How can the ISO 23247 standards be implemented in a Digital Twin lab setting, integrated with sustainability, AI, and XR technologies for Industry 5.0?*

First, we discuss the components of the DT framework and analyze how they align with the ISO 23247 standard, highlighting the standard's role in structuring our implementation. Second, we examine the architecture of the DTs in relation to existing literature, comparing our approach with established models to identify advancements and contributions. Third, we explore the interrelated elements within the DTs framework, illustrating how these components co-relate and work synergistically to enhance the system's overall functionality and value.

**FIGURE 6 FUNCTIONAL VIEW OF THE INSTANTIATED REFERENCE ARCHITECTURE**

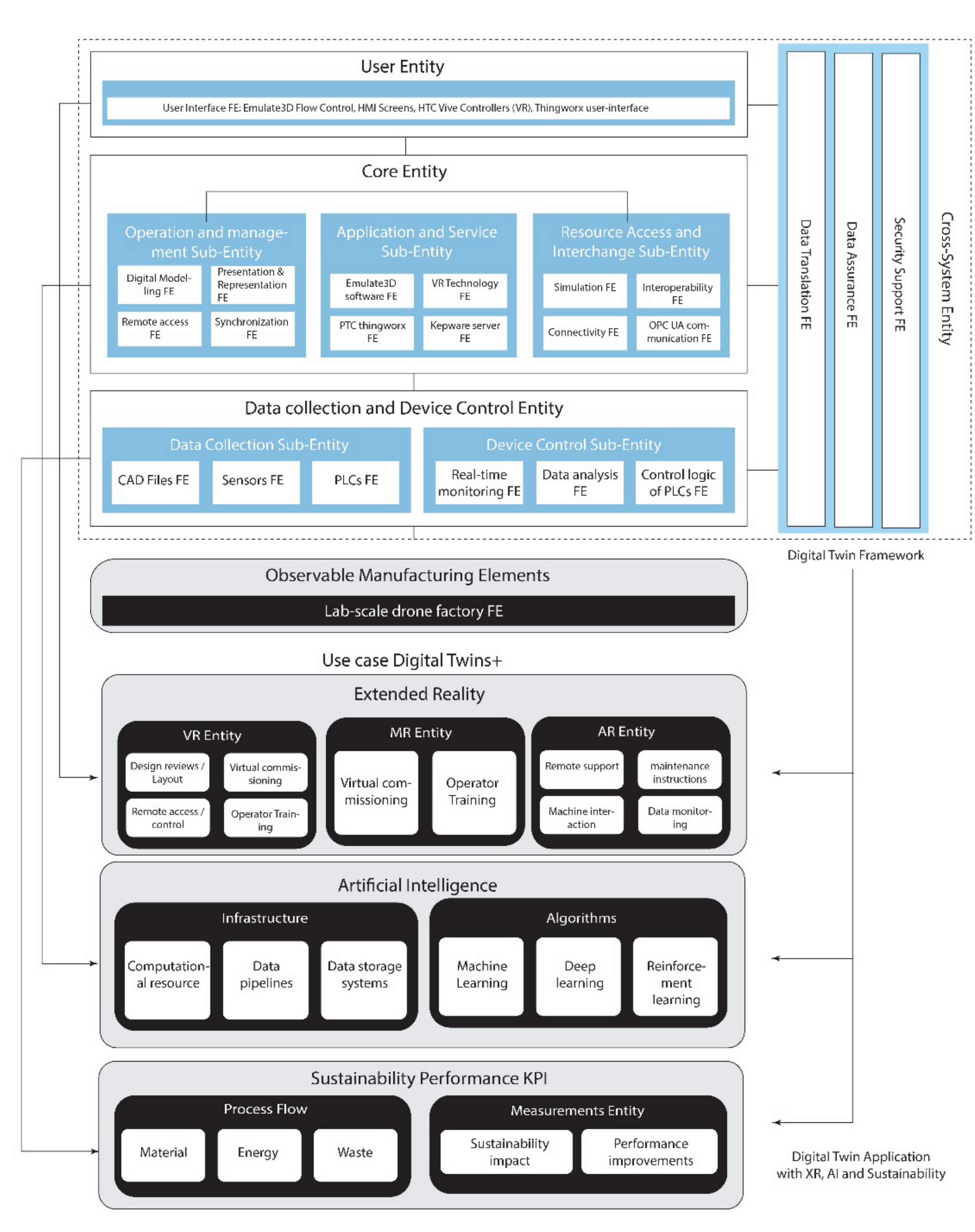

By systematically validating each aspect of the DT framework, we have demonstrated its potential to enhance manufacturing processes, as shown in Figure 6 based on the implementation of ISO23247. The integration of VR not only provides an immersive user experience but also opens avenues for future research into social sustainability aspects such as cognitive ergonomics, operator training, and improved decision-making support. Besides, AI and sustainability have been included to expand the core entity and data collection /device control entity. The validated DT framework sets a foundation for exploring advanced applications like AI-driven analytics, remote monitoring, and virtual commissioning, sustainability performance KPI, all contributing to sustainable and efficient production systems.

## 5.1 XR USE CASES AND BENEFITS

This paper demonstrates the proof-of-concept by establishing the use case of a real-time connected DT where communication and control are bi-directional, with future applications including remote control and monitoring with VR accessibility. The drone factory real-connectivity use case has shown the user entity of the framework to have VR as a user interface to do the remote monitoring and control through the DTs. Moreover, VR can be expanded to multiple use cases that bring the potential of the DTs to wider use cases and applications. VR is integrated into the system through the simulation software that supplies both the PLC protocol and VR streaming function and has been regarded as the key to making the VR-integrated DT successful. By immersing users into the virtual side of the DTs and enabling real interaction through VR, we can anticipate a number of use cases where the combination of DTs technology and VR shows potential benefits.

### Virtual training

The use of VR for operator training has long been discussed both in literature and industry. The applications of VR are now ubiquitous in vocational training in various fields requiring hands-on practice such as production, military, and healthcare (Pawel et al., 2023). VR training allows users to get hands-on experience without interfering with an actual production system or the requirement for real equipment, enabling safe, remote, and efficient training environments (Setia Hermawati et al., 2015). Utilizing a DT for VR training is, however, a slightly less explored topic in literature but has the potential to increase the capabilities of VR training. Not only will the DTs offer a realistic virtual model or environment for training, but also realistic scenarios based on the real-time connectivity and data sharing from the physical system to the virtual system used for training (Kaarlela, Pieska, and Pitkaaho 2020). This enables greater insight into not only *how* we train but also *what* we train by utilizing real system data for realistic tasks and training creation (Alberto et al., 2024).

### Virtual design reviews

Using VR for collaborative meetings and design reviews has also been discussed frequently in literature and has seen application in industry for both product and production design (Schäfer et al., 2022). VR and XR devices bring increased capabilities of visualization and collaboration in the design process enabling multiple stakeholders to interact with and view the same problem for deeper discussions and evaluations (Garcia Rivera et al., 2024). DTs provide the capability to enhance the design reviews of both products and production by reviewing a concept design in a virtual production system operation. Real-time data and realistic representation of the production enable the stakeholders to visualize and analyze the product in a real production scenario based on collected operational data. For instance, understanding of how a production disturbance would affect a particular design change to a machine or how the assembly ergonomics on the intended workstation is affected by a new product design change. DTs of stations can minimize the time to develop and design an assembly line, preventing ergonomic problems and correcting design errors during the design phase (Caputo et al., 2019).

**Virtual commissioning and operation**

The topic of virtual commissioning is also an area where DT and VR technologies have the potential to introduce improved interfaces in real-time (Kuts, V. et al., 2021). For instance, DTs with VR enable the commissioning and reprograming of a robot in real-time, remotely, using improved human programing interfaces, such as manual jogging or teaching (Mashur, C. et al., 2020). Furthermore, data-based simulations and scenario analysis performed by the DT can be visualized using VR, improving our ability to comprehend various future scenarios for decision-making in our commissioning or day-to-day operations.

In this paper, we do not explore the use of any XR technologies as an interface to the real system and physical side of the DTs. However, by consulting the virtual continuum (Milgram et al., 1994) the conclusion can be drawn that AR or MR could be seen as potential technologies to facilitate such an interface. By utilizing AR or MR technologies, interactions with the physical assets and system could have been achieved by overlaying data and information from the DTs. In such a scenario we could anticipate additional use cases where the combination of the DT technology and AR/MR shows potential benefits.

**Data monitoring and feedback**

Using AR or MR technologies, production or sensor data can be overlayed directly on a workstation or machine for operators to quickly and easily identify deviations, problems, and information that needs their attention (Zhu et al., 2019). Furthermore, by combining this with a DT, we could take it one step further by visualizing simulated and anticipated data from the simulation model directly to the operator within the production system to act fast to the new information and anticipated scenarios (Zhu et al., 2019).

**Operator cognitive support and instructions**

Similar to the previous use case, AR and MR technologies could also be used to directly provide instructions and support operators to present in the physical production system, often in terms of assembly-machine- or work instructions (Yin, Y. et al., 2023). Such instruction could be supplied directly from the DT and simulation model to the operator adapted to the real-time data and scenario captured by the DT (Yin, Y. et al., 2023). This would enable flexible and efficient instructions to be present at the right time and the right place. Such instructions could be adapted after the current state of production or even anticipated future scenarios to prevent undesired production disturbances.

## 5.2 AI INTEGRATED CORE ENTITY

The established connectivity can be used to control both the simulation model and the real system in the drone factory use case. AI can potentially expand the core entity of the DTs framework from the lifecycle perspective. It can enhance resource accessibility and operation management with expanded use cases. Not only can it easily be applied to the design phase, as in virtual commissioning, but also to the maintenance phase. Implementing AI analysis is beneficial as it aids in evaluating machine health, e.g., preventive maintenance, or exploration of what-if scenarios (Taylor et al., 2021).

With the advancement of AI, particularly emerging technologies like generative AI, DTs are becoming more capable of addressing current challenges. They can now simulate with greater accuracy, enhance predictive maintenance, and even generate hypothetical scenarios (Lv et al., 2022). However, in our drone assembly DT model, the AI techniques are still not applied to support the decision-making and analysis. Regarding the value and benefits of AI-supported DTs, two general aspects can be considered: infrastructure and algorithms.

Infrastructure is the foundation for applying AI to DTs (Chen et al., 2023). In the DT model, the data storage system is critical to managing the large amount of data generated by sensors and other sources. These systems must be designed to efficiently and securely store both structured and unstructured data, enabling fast access and processing. In addition, computational resources, including high-performance computing and cloud processing power, allow for the incorporation of complex AI techniques into the model. Data pipelines further enhance these infrastructures, enabling smooth data flow between components, pipeline data preprocessing, and ensuring quality and consistency, ultimately providing a solid foundation for AI applications in DT.

Algorithms are central to supporting the analysis and decision-making capabilities of AI-driven DTs. Machine learning techniques enable DTs to recognize patterns in historical data and enhance predictive maintenance and anomaly detection (Chen et al., 2023). For example, machine learning can detect anomalous problems in the drone assembly line and identify a potential problem before it appears or escalates. Deep learning, a subset of machine learning, brings more sophisticated pattern recognition and data processing capabilities to its neural networks, enabling DTs to process large amounts of complex, high-dimensional data for further advanced analyses. Reinforcement learning takes this a step further, enabling DTs to learn and optimize decision-making strategies through continuous interaction and feedback from simulated environments, as well as reinforcement learning's ability to use DTs themselves as training environments to explore different scenarios and outcomes (Cronrath et al., 2019; Dai et al., 2020). All of them can provide analytical capabilities to empower decision-making, supporting real-time decision-making, predictive analytics, and optimization in DT models.

Looking ahead, AI will drive DTs towards fully autonomous systems, enabling them not only to predict and simulate outcomes but also to optimize operations, make decisions, and take corrective actions in real-time. This reduces reliance on historical data, allowing DTs to anticipate and sense the state of physical systems proactively.

## 5.3 DATA COLLECTION AND CONTROL ENTITY FOLLOWING SUSTAINABILITY PERFORMANCE KPI

The study presented in this paper shows that people are gaining an increasing understanding that the use of digital technologies must take place under defined rules (Schuh et al., 2023). However, there is a large gap between hopes and reality when it comes to the environmental implications of digitalization (Santarius et al., 2023). A review study summarized three categories of 10 major demands for the role of DTs regarding the context of sustainable production: requirements regarding data interoperability, requirements regarding the data itself, and requirements regarding specific sub-models (Miehe et al., 2021). The architecture presented in Figure 6 reflects several of the demands, indicating potential connections between this standardized DT demonstration and sustainable production.

The DT applications could preliminarily aid the data collection of three process flows used mostly in sustainability performance analyses: material, energy, and waste (Smith & Ball, 2012). Furthermore, instead of manual work through unstandardized templates, DTs ensure the interoperability of relevant data and documents through harmonized formats within an adaptable structure.

In the measurement entity, sustainable performance measurement and variant analysis can be supported by directly connecting DTs and simulations relevant to environmental impact assessment (for example, the life cycle assessment) and predictive models for the improvement of sustainability performance.

## 5.4 FUTURE RESEARCH

Building upon the findings of the study, future research could branch in diverse directions. First is the user investigation with quantitative and qualitative data as support to validate the benefits of DTs and the demonstrations. Such efforts are essential to evaluate the benefits of DTs on social sustainability aspects, including cognitive ergonomics, enhanced decision-making support and pedagogical advancements in virtual training environments. While this study has showcased the feasibility of enabling DT use cases within specific simulation software and physical systems by adhering to the ISO 23247 standard, there remains a need to explore the full spectrum of benefits and stability of the solution and potential safety risks.

As for a general takeaway regarding the implementation of a DT for monitoring and control of manual assembly, the security issues and the purpose of the DT need to be investigated before development. For instance, a moving assembly line controlled remotely can risk the safety of operators if they are within the vicinity of the assembly line when a request is being executed. To ensure the safety of operators, pressure-sensitive mats can be installed blocking control of the assembly line remotely when an operator is standing on it. For this specific use case, the virtual platform should have strict authority access and should be operated as a monitoring system, emergency maintenance support, and remote-control system when no operators are working in the workstation area. Another concern regards security and privacy in the adoption of Industry 5.0. It is a critical challenge to establish robust security measures to foster trust within ecosystems that utilize interconnected devices. As reliance on artificial intelligence and automation increases, the potential for security threats grows, necessitating stringent security protocols. Solutions must be implemented to address vulnerabilities, including authentication and access control issues in IoT systems, to safeguard sensitive data against risks like man-in-the-middle attacks (Adel 2022).

Immersive user interfaces of DTs supported by AI will not only facilitate better understanding of real-time data, but also contribute to user well-being and efficiency in sustainable production systems. DTs will not only be a tool for automation, but also bridge the industry 4.0 to 5.0 with human factors into consideration in its scalability.

To explore how DT can be implemented as an enabler for environmental performance improvement, future use cases need to consider the data collection of key process flows as a part of DT infrastructure. A data inventory for environmental impact assessments can also be shared with other entities. For instance, the predictive model of environmental performance requires AI integrated core entity for enhancing data completeness; and the interface design for better environmental performance visualization are essential at workplaces.

## 6 Conclusion

The paper demonstrates how connectivity in DTs can potentially drive improvements in system operations and decision-making processes with real-time simulation, analysis, and monitoring, with VR integrated to highlight the human-centric perspective in the technology. It presents a reference architecture supporting the integration of IoT platforms with 3D simulation software based on the ISO 23247 DT Framework. The objective is to bridge the current gaps in pragmatic implementation strategies and the verification of frameworks, which was provided by the integration of the simulation model and the IoT platform.

To further scale the deployed DT, the IoT solution requires further development, such as several layers of Kepware servers to ensure secure communications throughout the organization, and additional handshakes in the communication, such as checking if a mission has been completed. For a fully synchronized DT, all sensors and actuators need to be connected, where in this case only selected sensors belonging to specific push buttons are connected. Together with applying a purpose to the DT, for instance, analysis supported by AI of current and future systems will bring business value to the DT and will be the matter for future research.

To better understand how to realize the benefits of real-time connectivity, further studies with concrete use cases should be designed and explored with multiple stakeholders, including operators, engineers, and managers. However, regardless of the risk, implementing DTs correctly in the factory will still be the key to accelerate the digital transformation of the industry. Investigating the implications of integrating AI and VR user interfaces into DTs can provide deeper insights into social sustainability aspects like cognitive ergonomics, and also bring more use cases to enable sustainable production with visualization to bridge DTs for Industry 5.0. However, exploring security measures and establishing best practices in authorization will be vital in mitigating risks associated with remote control and monitoring capabilities. Future progression on AI enhanced sustainability and human centric XR for DTs will not only promote sustainable and efficient production systems with DTs system but also human-centric consideration, ensuring the technology advancement aligns with the well-being and effectiveness of its users and stakeholders to align with Industry 5.0 agenda.

## ACKNOWLEDGMENTS

The authors would like to thank the Swedish innovation agency VINNOVA for their funding of the PLENUM project, grant number: 2022-01704 and Digitala Stambanan, grant number: 2021-02421. The work was carried out within Chalmers's Area of Advance Production. The support is gratefully acknowledged.

## DISCLAIMER

Specific commercial products and systems are identified in this paper to facilitate understanding. Such identification does not imply that these software systems are necessarily the best available for the purpose. No approval or endorsement of any commercial product by NIST is intended or implied.

## REFERENCES

Adel, A. 2022. "Future of Industry 5.0 in Society: Human-Centric Solutions, Challenges, and Prospective Research Areas." *Journal of Cloud Computing: Advances, Systems and Applications* 11, no. 1: Article 29. https://doi.org/10.1186/s13677-022-00314-5.

Martínez-Gutiérrez, Alberto, Javier Díez-González, Hilde Perez, and Madalena Araújo. Forthcoming. "Towards Industry 5.0 through Metaverse." *Robotics and Computer-Integrated Manufacturing*. https://doi.org/10.1016/j.rcim.2024.102764.

Ariansyah, D., I. Fernández del Amo Blanco, J. Erkoyuncu, M. Agha, D. Bulka, and J. Puante. 2020. "Digital Twin Development: A Step by Step Guideline." *SSRN Electronic Journal.,* https://doi.org/10.2139/ssrn.3717726.

Asad, U., M. Khan, A. Khalid, and W. A. Lughmani. 2023. "Human-Centric Digital Twins in Industry: A Comprehensive Review of Enabling Technologies and Implementation Strategies." *Sensors 23*, no. 8: 3938. https://doi.org/10.3390/s23083938.

Breque, M., de Nul, L., Petridis, A. 2021. "Industry 5.0 - towards a sustainable, human-centric and resilient European industry". *Policy Brief European Commission*

Cao, H., H. Söderlund, M. Despeisse, F. Rivera, and B. Johansson. 2024. "VR Interaction for Efficient Virtual Manufacturing: Mini Map for Multi-User VR Navigation Platform." *Advances in Transdisciplinary Engineering*. https://doi.org/10.3233/ATDE240178.

Caputo, F., A. Greco, M. Fera, and R. Macchiaroli. 2019. "Digital Twins to Enhance the Integration of Ergonomics in the Workplace Design." *International Journal of Industrial Ergonomics* 71: 20–31. https://doi.org/10.1016/j.ergon.2019.02.001.

Cronrath, C., A. R. Aderiani, and B. Lennartson. 2019. "Enhancing Digital Twins Through Reinforcement Learning." In *Proceedings of the 2019 IEEE 15th International Conference on Automation Science and Engineering (CASE)*, 293–298. New York: IEEE. https://doi.org/10.1109/CASE.2019.8843244.

Chen, S., J. P. G. Sánchez, E. T. Bekar, J. Bokrantz, A. Skoogh, and P. V. Lopes. 2024. "Understanding Stakeholder Requirements for Digital Twins in Manufacturing Maintenance." In *Proceedings of the Winter Simulation Conference (WSC '23),* 2008–2019. New York: IEEE Press.

Chen, C., H. Fu, Y. Zheng, F. Tao, and Y. Liu. 2023. "The Advance of Digital Twin for Predictive Maintenance: The Role and Function of Machine Learning." *Journal of Manufacturing Systems* 71: 581–594. https://doi.org/10.1016/j.jmsy.2022.12.009.

Dai, Y., K. Zhang, S. Maharjan, and Y. Zhang. 2020. "Deep Reinforcement Learning for Stochastic Computation Offloading in Digital Twin Networks." *IEEE Transactions on Industrial Informatics 17*, no. 7: 4968–4977.

Da Silva, A. G., M. V. M. Gomes, and I. Winkler. 2022. "Virtual Reality and Digital Human Modeling for Ergonomic Assessment in Industrial Product Development: A Patent and Literature Review." *Applied Sciences 12*, no. 3: 1084. https://doi.org/10.3390/app12031084.

IBM Cloud. 2024. "IBM Watson Internet of Things." Accessed October 2, 2024. https://internetofthings.ibmcloud.com/.

Domínguez-Bolaño, T., O. Campos, V. Barral, C. J. Escudero, and J. A. García-Naya. 2022. "An Overview of IoT Architectures, Technologies, and Existing Open-Source Projects." *Internet of Things 20*: Article 100626. https://doi.org/10.1016/j.iot.2022.100626.

Emmert-Streib, F. 2023. "What Is the Role of AI for Digital Twins?" *AI 4*, no. 3: 721–728.

Garcia Rivera, F., A. Rostami, S. Mattsson, and H. Söderlund. 2024. "How Can XR Enhance Collaboration with CAD/CAE Tools in Remote Design Reviews?" *Advances in Transdisciplinary Engineering*. https://doi.org/10.3233/ATDE240182.

Fan, W., X. Guo, E. Feng, J. Lin, Y. Wang, J. Liang, M. Garrad, J. Rossiter, Z. Zhang, N. Lepora, L. Wei, and D. Zhang. 2023. "Digital Twin-Driven Mixed Reality Framework for Immersive Teleoperation with Haptic Rendering." *IEEE Robotics and Automation Letters 8*, no. 12: 8494–8501. https://doi.org/10.1109/LRA.2023.3325784.

Farshid, M., Paaschen, J., Eriksson, T., Kietzmann, J., 2018, "Go boldly!: Explore augmented reality (AR), virtual reality (VR), and mixed reality (MR) for business", *Business Horizons*, https://doi.org/10.1016/j.bushor.2018.05.009.

George, Henry. 2017. "Henry George, Letter, October 19, 1889, to Hamlin Garland." https://doi.org/10.25549/GAR-C81-12347.

Greco, A., M. Caterino, M. Fera, and S. Gerbino. 2020. "Digital Twin for Monitoring Ergonomics During Manufacturing Production." *Applied Sciences 10*, no. 21: 7758. https://doi.org/10.3390/app10217758.

Grieves, M. 2014. "Digital Twin: Manufacturing Excellence Through Virtual Factory Replication." *White Paper 1,* no. 2014: 1–7.

Grieves, M., and J. Vickers. 2017. "Digital Twin: Mitigating Unpredictable, Undesirable Emergent Behavior in Complex Systems." In *Transdisciplinary Perspectives on Complex Systems: New Findings and Approaches,* edited by Editors' Names, 85–113. Springer. https://doi.org/10.1007/978-981-10-5861-5_4.

Guth, J., U. Breitenbücher, M. Falkenthal, P. Fremantle, O. Kopp, and F. Leymann. 2018. "A Detailed Analysis of IoT Platform Architectures: Concepts, Similarities, and Differences." In *Transdisciplinary Perspectives on Complex Systems: New Findings and Approaches,* 81–101. Springer. https://doi.org/10.1007/978-981-10-5861-5_4.

Guth, J., U. Breitenbücher, M. Falkenthal, F. Leymann, and L. Reinfurt. 2016. "Comparison of IoT Platform Architectures: A Field Study Based on a Reference Architecture." In *2016 Cloudification of the Internet of Things (CIoT),* 1–6. New York: IEEE. https://doi.org/10.1109/CIOT.2016.7872918.

Havard, V., B. Jeanne, M. Lacomblez, and D. Baudry. 2019. "Digital Twin and Virtual Reality: A Co-Simulation Environment for Design and Assessment of Industrial Workstations." *Production & Manufacturing Research 7*, no. 1: 472–489. https://doi.org/10.1080/21693277.2019.1660283.

IEEE. 2020. IEEE Standard for an Architectural Framework for the Internet of Things (IoT). *IEEE Std 2413-2019: 1–269*. https://doi.org/10.1109/IEEESTD.2020.9032420.

Jazdi, N., B. A. Talkhestani, B. Maschler, and M. Weyrich. 2021. "Realization of AI-Enhanced Industrial Automation Systems Using Intelligent Digital Twins." *Procedia CIRP* 97: 396–400.

Jiang, Y., S. Yin, K. Li, H. Luo, and O. Kaynak. 2021. "Industrial Applications of Digital Twins." *Philosophical Transactions of the Royal Society A: Mathematical, Physical and Engineering Sciences 379*, no. 2207: Article 20200360. https://doi.org/10.1098/rsta.2020.0360.

Kaarlela, T., S. Pieska, and T. Pitkaaho. 2020. "Digital Twin and Virtual Reality for Safety Training." *2020 IEEE 11th International Conference on Cognitive Infocommunications (CogInfoCom),* 115–120. https://doi.org/10.1109/CogInfoCom50765.2020.9237812.

Kim, D. B., G. Shao, and G. Jo. 2022. "A Digital Twin Implementation Architecture for Wire + Arc Additive Manufacturing Based on ISO 23247." *Manufacturing Letters* 34: 1–5.

Kuts, V., et al., 2021. "Digital Twin: Universal User Interface for Online Management of the Manufacturing System." *Advanced Manufacturing.* https://doi.org/10.1115/IMECE2021-69092.

Liao, Y., H.-Y. Tseng, Y.-J. Lin, C.-J. Wang, and W. Hsu. 2020. "Using Virtual Reality-Based Training to Improve Cognitive Function, Instrumental Activities of Daily Living and Neural Efficiency in Older Adults with Mild Cognitive Impairment: A Randomized Controlled Trial.," *European Journal of Physical and Rehabilitation Medicine.* https://doi.org/10.23736/S1973-9087.19.05899-4.

Lierop, M., V. Allard, and J. Hurk. 2021. "Augmented Reality." In *Industry, Innovation and Infrastructure*. https://doi.org/10.1007/978-3-319-95873-6_300006.

Lopes, P. V., S. Chen, J. P. G. Sánchez, E. T. Bekar, J. Bokrantz, and A. Skoogh. 2023. "Data-Driven Smart Maintenance Decision Analysis: A Drone Factory Demonstrator Combining Digital Twins and Adapted AHP." In *2023 Winter Simulation Conference (WSC),* 1996–2007. https://doi.org/10.1109/WSC60868.2023.10408351.

Lv, Z., and S. Xie. 2022. "Artificial Intelligence in the Digital Twins: State of the Art, Challenges, and Future Research Topics." *Digital Twin* 1: 12.

Lugaresi, G., Z. Jemai, and E. Sahin. 2023. "Digital Twins for Supply Chains: Main Functions, Existing Applications, and Research Opportunities." In *2023 Winter Simulation Conference (WSC)*, 2076–2087. IEEE.

Maruyama, T., T. Ueshiba, M. Tada, H. Toda, Y. Endo, Y. Domae, Y. Nakabo, T. Mori, and K. Suita. 2021. "Digital Twin-Driven Human Robot Collaboration Using a Digital Human." *Sensors* 21, no. 24: 8266. https://doi.org/10.3390/s21248266.

Mashur, C., K. Konkol, A. Geiger, and R. Stark. 2020. "VR-Based Development Tools for Virtual Commissioning Using Digital Twins." *ZWF Zeitschrift fuer Wirtschaftlichen Fabrikbetrieb*. https://doi.org/10.3139/104.112411.

Miehe, R., L. Waltersmann, A. Sauer, and T. Bauernhansl. 2021. "Sustainable Production and the Role of Digital Twins—Basic Reflections and Perspectives." *Journal of Advanced Manufacturing and Processing* 3, no. 2: e10078. https://doi.org/10.1002/amp2.10078.

Milgram, P., and F. Kishino. 1994. "Taxonomy of Mixed Reality Visual Displays." *IEICE Transactions on Information and Systems* E77-D, no. 12: 1321–1329.

Neumann, W. P., Winkelhaus, S., Grosse, E. H., & Glock, C. H. 2020b. "Industry 4.0 and the human factor – A systems framework and analysis methodology for successful development". *International Journal of Production Economics*, 233, 107992. https://doi.org/10.1016/j.ijpe.2020.107992

Ottogalli, Rosquete, Amundarain, Aguinaga, and Borro. 2019. "Flexible Framework to Model Industry 4.0 Processes for Virtual Simulators." *Applied Sciences* 9, no. 23: 4983. https://doi.org/10.3390/app9234983.

Strojny, Paweł, and Natalia Dużmańska-Misiarczyk. 2023. "Measuring the Effectiveness of Virtual Training: A Systematic Review." *Computers & Education: X Reality*. https://doi.org/10.1016/j.cexr.2022.100006.

Pavlou, M., D. Laskos, E. I. Zacharaki, K. Risvas, and K. Moustakas. 2021. "XRSISE: An XR Training System for Interactive Simulation and Ergonomics Assessment." *Frontiers in Virtual Reality* 2. https://doi.org/10.3389/frvir.2021.646415.

Parrott, Aaron, and L. Warshaw. 2017. "Industry 4.0 and the Digital Twin: Manufacturing Meets Its Match." *Deloitte Insights*. https://www2.deloitte.com/us/en/insights/focus/industry-4-0/digital-twin-technology-smart-factory.html.

Perno, M., L. Hvam, and A. Haug. 2022. "Implementation of Digital Twins in the Process Industry: A Systematic Literature Review of Enablers and Barriers." *Computers in Industry* 134: Article 103558. https://doi.org/10.1016/j.compind.2021.103558.

Pirker, J., E. Loria, S. Safikhani, A. Künz and S. Rosmann. 2022. "Immersive virtual reality for virtual and digital twins: A literature review to identify state of the art and perspectives". In *2022 IEEE Conference on Virtual Reality and 3D User Interfaces Abstracts and Workshops (VRW)*, 114–115. IEEE.

PTC Community Management. 2022. "Design Your Data Model Guide." Accessed April 23, 2024.

PTC Community Management. 2023. "Create an Application Key Guide." Accessed April 26, 2024.

Qi, Q., and F. Tao. 2018. "Digital Twin and Big Data Towards Smart Manufacturing and Industry 4.0: 360 Degree Comparison." *IEEE* Access 6: 3585–3593.

Rossini, R., D. Conzon, G. Prato, C. Pastrone, J. P. C. dos Reis, and G. Gonçalves. 2020. "REPLICA: A Solution for Next Generation IoT and Digital Twin-Based Fault Diagnosis and Predictive Maintenance." SAM IoT 2739: 55–62.

Saju, N. S., N. A. Babu, N. A. S. Kumar, N. T. John, and N. T. Varghese. 2022. "Augmented Reality vs Virtual Reality." *International Journal of Engineering Technology and Management Sciences* 6, no. 5: 379–383. https://doi.org/10.46647/ijetms.2022.v06i05.057.

Santarius, T., L. Dencik, T. Diez, H. Ferreboeuf, P. Jankowski, S. Hankey, A. Hilbeck, L. M. Hilty, M. Höjer, D. Kleine, S. Lange, J. Pohl, L. Reisch, M. Ryghaug, T. Schwanen, and P. Staab. 2023. "Digitalization and Sustainability: A Call for a Digital Green Deal.," *Environmental Science & Policy* 147: 11–14. https://doi.org/10.1016/j.envsci.2023.04.020.

Schäfer, A., G. Reis, and D. Stricker. 2022. "A Survey on Synchronous Augmented, Virtual, and Mixed Reality Remote Collaboration Systems." *ACM Computing Surveys*. https://doi.org/10.1145/3533376.

Schuh, G., M.-F. Stroh, and L. Johanning. 2023. "Toward Responsible Use of Digital Technologies in Manufacturing Companies Through Regulation." In *Proceedings of the Conference on Production Systems and Logistics:* CPSL 2023 - 1. Hannover : publish-Ing., 2023, S. 280-289. DOI: https://doi.org/10.15488/13447

Schuh, G., R. Anderl, R. Dumitrescu, A. Krüger, and M. ten Hompel. 2020. "Industrie 4.0 Maturity Index: Managing the Digital Transformation of Companies." Accessed March 19, 2024.

Setia Hermawati, et al., 2015. "Understanding the Complex Needs of Automotive Training at Final Assembly Lines." *Applied Ergonomics* 46: 201–215. https://doi.org/10.1016/j.apergo.2014.07.014.

Simões, B., R. De Amicis, I. Barandiaran, and J. Posada. 2019. "Cross Reality to Enhance Worker Cognition in Industrial Assembly Operations." *The International Journal of Advanced Manufacturing Technology* 105, no. 9: 3965–3978. https://doi.org/10.1007/s00170-019-03939-0.

Shao, G. 2021. "Use Case Scenarios for Digital Twin Implementation Based on ISO 23247." *Technical Report*. https://doi.org/10.6028/nist.ams.400-2.

Shao, G., S. Frechette, and V. Srinivasan. 2023. "An Analysis of the New ISO 23247 Series of Standards on Digital Twin Framework for Manufacturing." In *International Manufacturing Science and Engineering Conference*, Vol. 87240, V002T07A001. American Society of Mechanical Engineers.

Shao, G., and M. Helu. 2020. "Framework for a Digital Twin in Manufacturing: Scope and Requirements." *Manufacturing Letters* 24: 105–107. https://doi.org/10.1016/j.mfglet.2020.04.004.

Shao, G., J. Hightower, and W. Schindel. 2023. "Credibility Consideration for Digital Twins in Manufacturing." *Manufacturing Letters* 35: 24–28.

Sharma, A., E. Kosasih, J. Zhang, A. Brintrup, and A. Calinescu. 2022. "Digital Twins: State of the Art Theory and Practice, Challenges, and Open Research Questions." *Journal of Industrial Information Integration* 30: Article 100383. https://doi.org/10.1016/j.jii.2022.100383.

Shen, Z., F. Arraño-Vargas, and G. Konstantinou. 2023. "Artificial Intelligence and Digital Twins in Power Systems: Trends, Synergies, and Opportunities." *Digital Twin* 2, no. 11: Article 11.

Stena Industry Innovation Laboratory. 2024. "SII-lab - Opportunities for Industry and Society." Accessed January 20, 2024.

Shin, Y., and G. J. Kim. 2022. "XR-Based Interaction: Leveraging Virtual Digital Twin for Efficient Exploration with Small FOV Augmented Reality Glass." In *2022 IEEE International Symposium on Mixed and Augmented Reality Adjunct (ISMAR-Adjunct),* 808–809. https://doi.org/10.1109/ismar-adjunct57072.2022.00173.

Smith, L., & Ball, P. (2012). Steps towards sustainable manufacturing through modelling material, energy and waste flows. *International Journal of Production Economics*, 140(1), 227–238. https://doi.org/10.1016/j.ijpe.2012.01.036

Sun, W., W. Ma, Y. Zhou, and Y. Zhang. 2022. "An Introduction to Digital Twin Standards." *GetMobile: Mobile Computing and Communications* 26, no. 3: 16–22.

Swamy, S. N., and S. R. Kota. 2020. "An Empirical Study on System-Level Aspects of Internet of Things (IoT)." *IEEE* Access 8: 188082–188134. https://doi.org/10.1109/ACCESS.2020.3029847.

Tao, F., and M. Zhang. 2017. "Digital Twin Shop-Floor: A New Shop-Floor Paradigm Towards Smart Manufacturing." *IEEE* Access 5: 20418–20427.

Taylor, S. J. E., B. Johansson, S. Jeon, L. H. Lee, P. Lendermann, and G. Shao. 2021. "Using Simulation and Digital Twins to Innovate: Are We Getting Smarter?" In *2021 Winter Simulation Conference (WSC)*, 1–13. https://doi.org/10.1109/WSC52266.2021.9715535.

Tekinerdogan, B., and C. Verdouw. 2020. "Systems Architecture Design Pattern Catalog for Developing Digital Twins." *Sensors* 20, no. 18: 5103. https://doi.org/10.3390/s20185103.

TensorFlow. 2024. "TensorFlow Cloud." Accessed October 2, 2024. https://www.tensorflow.org/cloud?hl.

Wallner, B., B. Zwölfer, T. Trautner, and F. Bleicher. 2023. "Digital Twin Development and Operation of a Flexible Manufacturing Cell Using ISO 23247." *Procedia CIRP* 120: 1149–1154.

Xia, J., T. Jin, and L. Fan. 2022. "DTMR: Industrial Digital Twin Architecture for New Production Line Design in Mixed Reality Environment." In *2022 14th International Conference on Signal Processing Systems (ICSPS)*, 828–836. https://doi.org/10.1109/icsps58776.2022.00149.

Yang, C., J. Zhang, Y. Hu, X. Yang, M. Chen, M. Shan, et al., 2024. "The Impact of Virtual Reality on Practical Skills for Students in Science and Engineering Education: A Meta-Analysis." *International Journal of STEM Education* 11, no. 1. https://doi.org/10.1186/s40594-024-00487-2.

Yin, Y., Zheng, P., Li, X., Wang, L., 2023, "A state-of-the-art survey on Augmented Reality-assisted Digital Twin for futuristic human-centric industry transformation", *Robotics and Computer-Integrated Manufacturing*, 81, https://doi.org/10.1016/j.rcim.2022.102515.

Yuchen, J., Y. Shen, L. Kuan, L. Hao, and K. Okyay. 2021. "Industrial Applications of Digital Twins." *Philosophical Transactions of the Royal Society A: Mathematical, Physical and Engineering Sciences* 379, no. 2207: Article 20200360. https://doi.org/10.1098/rsta.2020.0360.

Zhang, G., C. Huo, L. Zheng, and X. Li. 2020. "An Architecture Based on Digital Twins for Smart Power Distribution System." In *2020 3rd International Conference on Artificial*

*Intelligence and Big Data (ICAIBD),* 29–33. IEEE. https://doi.org/10.1109/ICAIBD.2020.9182683.

Zhu, Z., Liu, C., Xu, X., 2019, " Visualisation of the Digital Twin data in manufacturing by using Augmented Reality", *Procedia CIRP*, 898-903, https://doi.org/10.1016/j.procir.2019.03.223.